\documentclass[published]{JHEP3} % 10pt is ignored!

\JHEP{00(2009)000}

\JHEPspecialurl{http://jhep.sissa.it/JOURNAL/JHEP3.tar.gz}

\usepackage{epsfig,multicol,bbm}

\newcommand\fverb{\setbox\fverbbox=\hbox\bgroup\verb}
\newcommand\fverbdo{\egroup\medskip\noindent%
			\fbox{\unhbox\fverbbox}\ }
\newcommand\fverbit{\egroup\item[\fbox{\unhbox\fverbbox}]}
\newbox\fverbbox

\def\gsim{\mathrel{\rlap{\lower 4pt \hbox{\hskip 1pt $\sim$}}\raise 1pt
\hbox {$>$}}}
\def\lsim{\mathrel{\rlap{\lower 4pt \hbox{\hskip 1pt $\sim$}}\raise 1pt
\hbox {$<$}}}

\title{Early Black Hole Formation by Accretion of Gas and Dark Matter}

\author{Hideyuki Umeda$^{a}$,
Naoki Yoshida$^{b}$, Ken'ichi Nomoto$^{b}$, 
Sachiko Tsuruta$^{c}$, Mei Sasaki$^{a}$ and Takuya Ohkubo$^{a}$\\
$^a$Department of Astronomy, School of Science, University of Tokyo, Hongo
Tokyo 113-0033, Japan\\
$^{b}$Institute for the Physics and Mathematics of the Universe, University of Tokyo
Kashiwa, Chiba 277-8568, Japan\\
$^{c}$Department of Physics, Montana State University, Bozeman, MT 59717, U.S.A.\\
E-mail: \email{umeda@astron.s.u-tokyo.ac.jp}}

\received{\today} 		%%
\accepted{\today}		%% These are for published papers.

\abstract{
Recent discovery of luminous quasars at $z > 6$ has
posed a severe challenge to the theory of structure formation of
the universe. These quasars are thought to be powered by
supermassive black holes (SMBHs). However no consensus is yet to
be reached as to the origin and early formation mechanism of
massive SMBHs.
We propose a model in which intermediate-mass black holes (IMBHs)
with mass of $\sim 10^4 M_{\odot}$ are formed in early
dark matter halos.
We carry out detailed stellar evolution
calculations for the first generation stars including annihilation
energy of dark matter (DM) particles.
We show that very massive stars, as massive as
$10^4M_{\odot}$, can be formed in an early DM halo.
Such stars are extremely bright with Log $L/L_\odot \gsim 8.2$.
They will gravitationally collapse to form IMBHs.
These black holes could have seeded the formation of early SMBHs.
}


\begin{document} 

%\maketitle  IS IGNORED %%%%%%%%%%%

\section{Introduction}

Recently bright quasars at redshifts greater than six were
discovered by the Sloan Digital Sky Survey\cite{Fan}. These
quasars are thought to be powered by SMBHs whose mass exceeds
$\sim 10^9 M_\odot$\cite{Willot}. The implication is that such
SMBHs must be in place as early as when the age of the universe is
about 800 Myr. It is often argued that early generation of stars
might have left BH remnants that could be the seeds for
SMBHs\cite{Haiman,Madau}. According to one such viable
scenario\cite{Volonteri}, seed black holes (BHs) of $\sim$
200$M_\odot$ are formed at redshifts of about 24 and these seeds
grow to become SMBHs through merging and highly super-Eddington
accretion. Some authors \cite{Li,Tanaka,BL03,SBH09} argued that early SMBH
formation is possible without invoking such a highly efficient
accretion process, but then these seed BHs have to be formed very
early in the universe. One way to ease these constrains is that
the seed BHs are much more massive, such as 
IMBHs with $\sim 500 - 10^5M_\odot$, 
rather than stellar mass BHs. Here we explore such a possibility.

 Structure formation in the universe is largely driven by
gravitational forces exerting on dark matter (DM) which is a major
matter content. DM plays a crucial role particularly in primordial
star formation\cite{Bromm}. The standard model of early structure
formation suggests that the first stars are formed in small-mass
DM halos with mass of $\sim 10^{5-6} M_{\odot}$\cite{Tegmark,Y03}
when the age of the universe is less than a few hundred million
years old. In this model, star-forming gas clouds are embedded at
the center of a DM halo, and hence the formation and
evolution of first stars are expected to be affected by DM.

 Weakly-interacting massive particles (WIMPs) are popular candidates for
dark matter. For these particles to be the bulk of dark matter
as thermal relics, they must have a large pair-annihilation
cross-section of the order $\left< \sigma v \right>_{\rm annihilation}
\sim 10^{-26}$ cm$^3$ s$^{-1}$.
Hence such WIMPs are expected to self-annihilate in high density regions, 
e.g., at the centers of
DM halos, where they are converted into high-energy
photons and particles. 
Although the DM annihilation signatures have not been
confirmed, observations with the FERMI satellite may prove the
existence of WIMPs. For reviews of dark matter candidates, 
see \cite{Bertone}.

 The effect of DM annihilation on
cosmic structure formation has been studied in various contexts
\cite{Ripamonti}. Effective heating by DM
annihilation could halt the collapse of a pre-stellar 
gas cloud\cite{Spolyar}. Primordial stars can also capture
DM particles\cite{Spergel,Gould},
which will then produce a significant energy at
the center of stars. These stars
comprise a new category -- stars powered and sustained
by the DM annihilation energy,
sometimes called ``dark stars'' \cite{Spolyar,Taoso,Yoon}. 

 Most previous works studied dark stars with constant masses.
However, it is likely that first stars largely grow their mass
during the evolution through mass-accretion. Therefore,
in this paper we calculate the growth and evolution of 
primordial stars using several plausible accretion models. 

\section{Dark matter model and calculation method}

 We consider the DM
capture via off-scattering\cite{Gould} and include energy
generation by DM annihilation in the stellar core. 
Several authors, instead, have
investigated DM supplied by the adiabatic contraction (AC).
As the gas collapses into the star, DM particles are gravitationally
pulled along with it. According to \cite{S09}, this
adiabatically contracted DM inside an accreting first stars 
runs out relatively
soon (at around $M\sim 780M_\odot$).
Also the adiabatically contracted DM density profile is roughly flat
inside a star, while the captured DM is thermalized and
highly concentrated at the center.
Therefore, if DM capture is effective, or the DM-baryon
scattering cross section is sufficiently large,
captured DM heating is expected to dominate 
for the later evolutionary stages of an accreting star. Hence,
we only consider the captured DM in this paper.
We note that adiabatic contraction may significantly change the early
evolution before the Kelvin-Helmholtz contraction stage,
but the results for $M >> 100M_\odot$ are almost independent
of these ``initial'' conditions.

 For the DM
particle parameters, we adopt commonly used ``fiducial'' values; 
the DM particle mass, $m_\chi = 100$ GeV, the
DM-baryon elastic scattering cross section $\sigma =
10^{-38}$cm$^{-2}$ \cite{Angle}, 
the DM self-annihilation
cross-section $\left< \sigma v \right>_{\rm annihilation} =
3\times 10^{-26}$ cm$^3$ s$^{-1}$.
The DM density at the center of the halo is assumed to be
$\rho_\chi =10^{11}$ GeV cm$^{-3}$\cite{Iocco}. We note that these parameters,
most notably the DM density, is still quite uncertain.
Since the DM annihilation
produces energy at a rate per unit volume, $\epsilon_{\rm DM}
=\left< \sigma v \right>_{\rm annihilation} \rho_\chi2/m_\chi$,
larger cross-section, larger DM density and smaller DM mass 
generally result in larger energy generation.
As long as the product of these quantities is similar to 
our fiducial value, the DM annihilation causes essentially the
same effect to those described in the present paper. 
We show a few test results with enhanced annihilation rates
in Section 4. 

 We mention that the adopted cross-section is close to the upper limits for spin-dependent 
interactions off a proton, recently derived from direct detection 
experiments XENON 10 \cite{Angle}.
The adopted DM density is much larger than the average DM density in
an early dark halo, but the ambient DM density around an accreting
population III star can be as large as this or even
larger, due to the adiabatic contraction of DM\cite{Freese09}.
Recently, Natarajan, Tan \& O'Shea \cite{Nata09} performed three-dimensional 
cosmological simulations and found that the central dark matter density
reached $\rho_\chi \sim 10^{10}$ GeV cm$^{-3}$ when the gas density
is $n_{\rm H} \sim 10^{10} {\rm cm}^{-3}$ in one of their simulations. 
Since the gas further collapses over 10 orders of magnitude in density
to the formation of a protostar \cite{YOH08}, we expect that the
dark matter density also increases by adiabatic contraction
(see the extrapolated density profile in \cite{Nata09}).
While three-dimensional effects during
the cloud collapse, such as tidal torques and turbulent motions 
of the gas will likely prevent dark matter to contract further, 
the fact that Natarajan et al. found at least one case 
with $\rho_{\chi, {\rm center}} \sim 10^{10}$ GeV cm$^{-3}$ 
at an intermediate stage is encouraging.

  There are two time scales relevant for dark star
evolution\cite{IoccoB}. One is $\tau_\chi$, and the equilibrium
between the dark matter (DM) capture and annihilation is
established after this time scale. The other is the thermal time
scale $\tau_{th} = 4\pi m_\chi R_*^{7/2} / 3\sqrt{2G}\sigma_0
M_*^{3/2}$. Here $R_*$ and $M_*$ are stellar radius and mass,
respectively. The captured DM gets redistributed in the interior
of the star in $\tau_{th}$, reaching a thermal distribution
$n_\chi(r)=n_\chi^c \rm{exp} (-r^2/r_\chi^2)$, with $r_\chi =
\sqrt{3kT_c/2\pi G\rho_c m_\chi}$, where $T_c$ and $\rho_c$ are
the core temperature and density,
$n_\chi^c=C\tau_\chi/\pi^{3/2}r_\chi^3$ is the DM density at the
stellar center, and $C$ is the DM capture rate 
which is explicitly given in \cite{Gould}. Since we assume a
dark star in equilibrium, these two time scales should be smaller
than the time step of the evolutionary calculations. We have
confirmed that $\tau_\chi$ is always sufficiently small and also
$r_\chi$ is much smaller than the core radius.

 $\tau_{th}$ is also sufficiently small for most of the
time except early periods of the evolution.
$\tau_{th}$ is typically larger than the stellar
evolutionary time-scale before the Kelvin-Helmholtz contraction
stage\cite{S09}.
In this case, the annihilation energy may be
exponentially suppressed. We have investigated this early
suppression and confirmed that the results perfectly
converged after the Kelvin-Helmholtz contraction stage.
In our models, the results converged for $M \gsim 30M_\odot$
and hence this effect does not change our conclusions.

% When a star becomes a red giant, $\tau_{th}(R) \propto R_*^{7/2}$
%increases and may exceed the computational time step. In this
%case, we simply include the DM capture inside a radius 'r'
%satisfying $t_1 > \tau_{th}(r)$, where $t_1$ is evolution time
%scale. We have found that the amount of DM captured by the
%extended outer envelope is not so large, and thus the decrease of
%the capture rate with this procedure is not so large. After the
%He-burning stage, the evolutionary time scale becomes shorter and
%shorter, and the assumption of equilibrium should be invalid.
%However as shown in Fig.4, at this stage, the nuclear energy
%generation completely dominates the DM annihilation energy, and
%thus the DM energy is no longer important.
 
 The DM annihilation luminosity $L_{DM}$ and the DM capture rate $C$ are
roughly proportional, such that $L_{DM} \propto C \propto \sigma
\rho_\chi/m_\chi$ when a DM star is in equilibrium\cite{IoccoB}.
The adopted DM density is much larger than the average DM density in
a dark halo, but the ambient DM density around an accreting
population III star is expected to be as large as this or even
larger due to the adiabatic contraction of DM\cite{Freese09}.

 We adopt the method in \cite{Ohkubo06,Ohkubo09} to calculate stellar
evolution from $M \sim 1-1.5M_\odot$ up to the onset of carbon
burning. We show that after this stage further evolution to
core-collapse is very rapid.

Of critical importance in our evolution calculations are the gas
mass accretion rates. We first adopt constant accretion rates of
$\dot{M} = 10^{-4}-10^{-2} M_{\odot} {\rm yr}^{-1}$, which will
cover a plausible range for primordial star formation \cite{OP03}.
However, the analytic models based on self-similar collapse
solutions \cite{ON98,MT08} and cosmological simulations
\cite{BL04,Y06} suggest that accretion rates will decrease with
time (and hence with increasing mass). Therefore, we also adopt
several cases with these time-dependent accretion rates. Also,
since the radiative feedback can reduce accretion rates in the
late evolution phase, we consider this effect also, following
\cite{MT08} as described below in Sec.3.1.

\section{Dependence on the gas mass accretion rates}

  The stellar evolution for models with constant gas mass accretion
already shows several interesting features. Fig.~1 shows the
evolution of mass and radius for the accreting stars without
(upper panel) and with (lower panel) DM annihilation heating. 

\smallskip
%\FIGURE{\epsfig{file=jhepfig.eps,width=5cm} 
%        \caption[Example of figure]
%{Made with \tt\ttbs FIGURE.}
%	\label{myfigure}}
\FIGURE{\epsfig{file=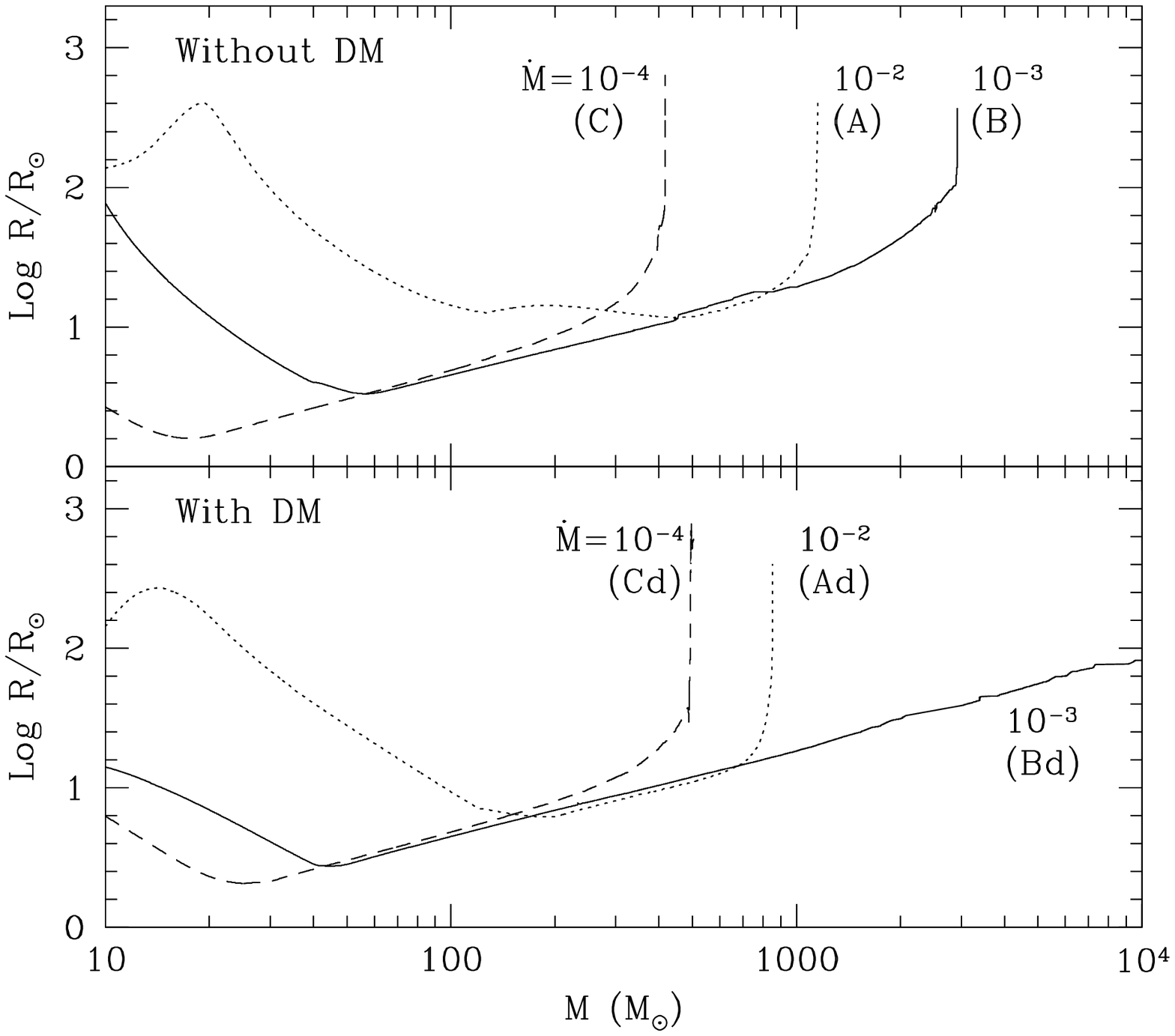,width=1\textwidth} 
        \caption
{Stellar mass - radius evolution for each model. The upper 
and lower panels show the models without and with the
DM annihilation energy, respectively. The model names for 
$\dot M=10^{-2}, 10^{-3}, 10^{-4} M_\odot$yr$^{-1}$
cases are A, B, C, and Ad, Bd, Cd, for the cases without and 
with the DM heating, respectively.}
	\label{fig1}}

\noindent
The adopted constant gas mass accretion rates are, $\dot{M} = 10^{-2}$
(Model A, Ad), $10^{-3}$ (B, Bd), and $10^{-4}$ (C, Cd) $M_\odot$
yr$^{-1}$, for without and with DM (for the letters without and
with the additional `d'). The overall evolution for the models
without DM heating is similar to those found in previous works
\cite{Ohkubo09,OP03}. With DM heating, all models go through the
Kelvin-Helmholz contraction phase when the stellar mass is several
to several tens solar masses, and eventually reach the
``main-sequence'' phase\cite{Ohkubo09}. 
Gravitational contraction is halted by
nuclear and/or DM annihilation energy generation. As Fig.2 shows,
the DM energy generation rate at the center exceeds the nuclear
energy generation for the models considered here. 

\smallskip
\FIGURE{\epsfig{file=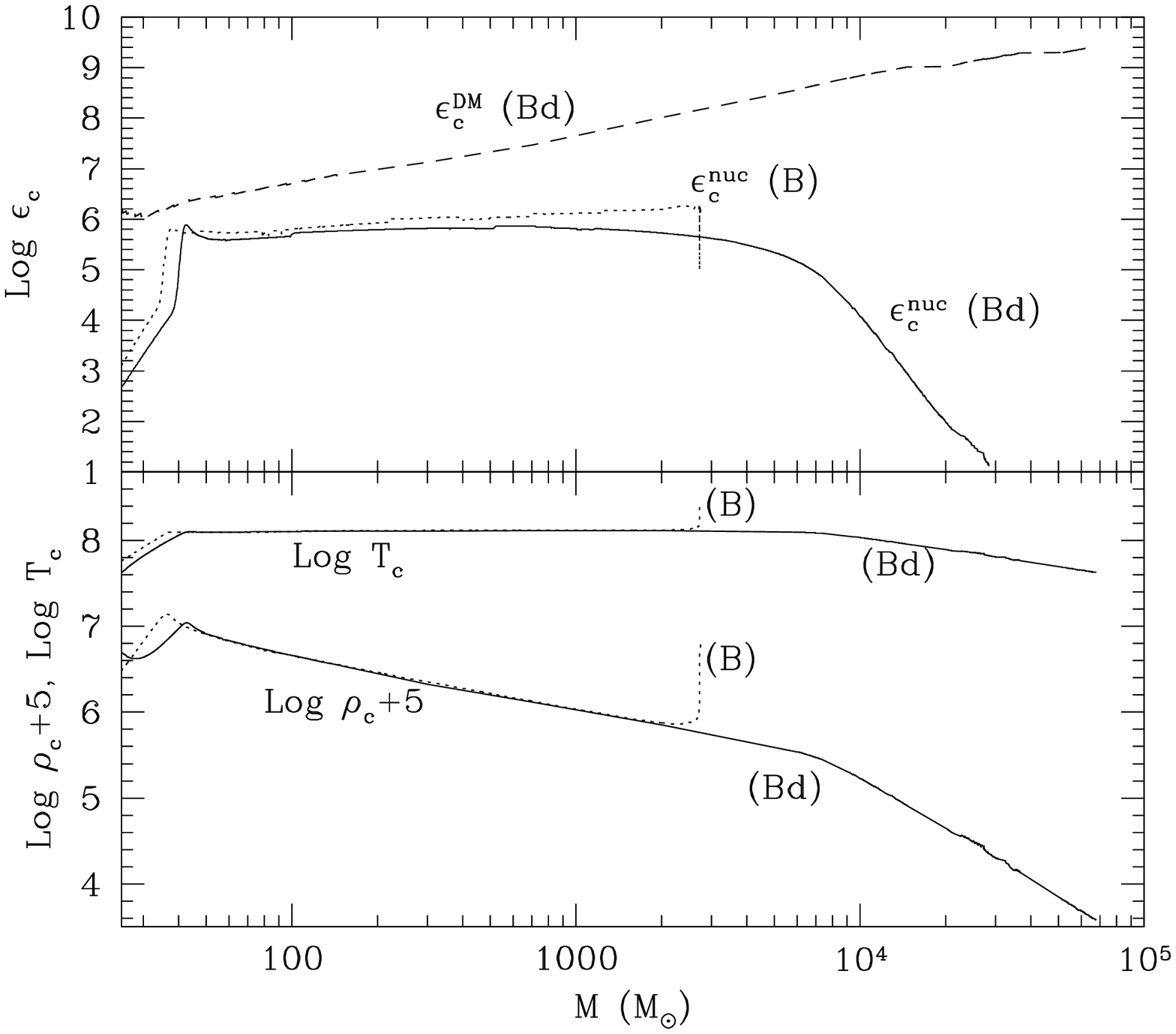,width=1\textwidth} 
        \caption{
Nuclear and dark matter annihilation energy generation
rate at center ($\epsilon_c$ in erg g$^{-1}$ s$^{-1}$)
(upper panel), and central density ($\rho_c$ in g cm$^{-3}$)
and temperature ($T_c$ in K)
(lower panel) as a function of stellar mass for models (B) and (Bd).
}\label{fig2}}

\noindent
Therefore, this star is a dark star and sustained  
by DM annihilation energy. However the
nuclear energy generation rate is not negligible; CNO-cycle
hydrogen burning is indeed taking place in the central part. The
stars evolve on the main-sequence track over a few million years,
with their luminosity increasing with mass. Fig.2 also shows the
central temperature and density as a function of stellar mass, for
B and Bd models.

 We find that the fate of the stars depends importantly on the
accretion rate. The DM annihilation supplies an extra energy to
support a star against gravitational contraction, and thus the
star consumes less hydrogen per unit time, with its lifetime
prolonged. The luminosity vs. ``final'' mass relation for 
various models are
shown in Fig. 3. Besides the steady accretion models A(d), B(d),
and C(d), various models with time-dependent accretion are also
shown. In many cases, the final stellar mass is
larger than 1000$M_\odot$.

\smallskip
\FIGURE{\epsfig{file=fig3.ps,width=1\textwidth} 
        \caption{
Final mass and
luminosity for various models. Models A(d), B(d), C(d) are the
same as in Fig.1. Models D, Y, F, and G are the cases with the
accretion rate $\dot M=10^{-5}M_\odot$yr$^{-1}$, rates in \cite{Y06},
\cite{MT08}, and \cite{Gao}, respectively. 
Lower case letters `d' in the model
names, such as Dd, indicate that the fiducial dark matter
annihilation energy is included. Models M and Md use the rates in
\cite{MT08} with a cut off at $M=300M_\odot$. In the models Cd$\times$3
and Fd$\times$3, the dark matter capture rate is enhanced by a
factor of 3 than the models Cd and Fd, respectively. The solid
straight line represents the Eddington Luminosity. The mass range
bounded by the dotted lines, labeled PISN, shows the mass range in
which the stars explode as pair-instability supernovae and do not
form massive BHs. The models with arrows are ``stalling'', or,
their central temperatures are decreasing. Therefore, their mass
may increase further.
}\label{fig3}}

 The most interesting is Model Bd, the case with
$\dot M = 10^{-3} M_\odot$yr$^{-1}$ (see Figs. 1, 2 and 3). It
lives long and continues to grow, and becomes a very massive star
with $M > 10^5 M_{\odot}$ (see Fig. 3). Hydrogen burning takes
place throughout the evolution after the zero-age main sequence,
but the DM heating becomes more and more important at later
evolutionary phases (see Fig. 2). When the stellar mass becomes 
$M \gsim 5000-6000 M_\odot$, the central convective regions where
hydrogen is partly exhausted, reaches the outer
hydrogen-unexhausted region, conveying fresh hydrogen into its
interior. Because of this convection, the central hydrogen
fraction starts to {\it increase} for $M \gsim 6000 M_\odot$, and
then the hydrogen consumption timescale ($\sim$ the star's
lifetime) becomes infinitely long, or the evolution is stalling.
The arrow for the model Bd in Fig.3 indicates that the model
is ``stalling'', or, its central temperature is decreasing. 
Therefore, its mass may increase further as long as the mass-accretion
continues with the same accretion rate.
%The final mass can be as large as $10^{5}M_\odot$ (see Fig. 3).
Note that without DM annihilation the final mass of this model (B)
is only modest, $\sim$ several thousand $M_{\odot}$.

 The stalling phase is expected to end, for example,
when the ambient DM density
becomes lower. The evolution after this stage toward the stellar
core-collapse is described in detail in Sec.5.

\subsection{Time dependent mass accretion rates}

 For the time-dependent mass accretion rates, we
use the ones given in \cite{Y06,Gao}. These authors performed
cosmological hydrodynamic simulations. They derive realistic gas
mass accretion rates for a typical population III stars formed at
around $z\sim 20$\cite{Y06}, and also for a very rare object
(``the very first star'') which forms around at $z\sim 50$ from a
peak of large-scale density fluctuations\cite{Gao}. Since the
$z\sim 6$ quasar is a rare object, using the rate in \cite{Gao}
is more suitable to explain the SMBH in the high redshifted
quasars. In Fig.3 we name the models using the rate
in \cite{Y06} and the rate of the model R5wt in \cite{Gao}, as
models Y and G, respectively. We use the following expression for
the accretion rate of model Y: $\dot M (M_\odot$yr$^{-1}) =
0.045(M/M_\odot)^{-2/3} ($for $M\leq 300M_\odot)$ and
$16.25(M/M_\odot)^{-1.7} ($for $M\geq 300M_\odot)$. For model G,
we use: $\dot M (M_\odot$yr$^{-1})$
 = min$(0.18(M/M_\odot)^{-0.6}, 6.0\times10^4 (M/M_\odot)^{-2.24})$.

 As mentioned above, the models G and Gd correspond to
very rare objects which form from peaks of
large-scale density fluctuations. In the hierarchical structure
formation model, such high density peaks grow eventually to
luminous quasars that host SMBHs at redshifts greater than
six\cite{Li,MR}. These models result in larger BH mass (several
thousands solar mass) than the models Y and Yd where the rate in
\cite{Y06} is used.

 These rates do not include possible radiative feedback effects from
the accreting star and accretion disk in the case of the
disk-like accretion.
Such effects for a disk-like accretion model was explored in \cite{MT08}.
In model F we assume the
same rate as model G for $M\lsim 90M_\odot$ but it reduces as $\dot M
=155.6 (M/M_\odot)^{-2.096}$ for larger masses. This rate is similar to the
model $K'=2$ in \cite{MT08}. In \cite{MT08}
it was suggested that stellar feedback terminates accretion
at the time when the disk evaporation time scale equals the
accretion time scale.
If this is the case, the final stellar mass would be upper bounded
as the model M in Fig.3.
However, these models themselves are also based on approximations and
geometrical simplifications. Since the star's lifetime is prolonged
in our model due to DM annihilation, longer timescale evolution
has to be followed in order to determine whether or not
gas accretion is completely quenched before the star dies.

 The final masses for the models F and Fd are both
$\sim 1000M_\odot$ and the effects of the DM annihilation is not so large
for the fiducial DM parameter set. However as shown in the next 
section if the DM annihilation energy is larger, the final mass
would be much larger as in the model Fd$\times 3$ in Fig.3.

% There are also several interesting models which are based on
%time-dependent accretion rates. 
%For instance, models G and Gd are
%based on the rate in \cite{Gao} from a very large-volume
%cosmological simulation by these authors. The simulation follows
%the formation of a very rare object which forms from a peak of
%large-scale density fluctuations. In the hierarchical structure
%formation model, such high density peaks grow eventually to
%luminous quasars that host SMBHs at redshifts greater than
%six\cite{Li,MR}. These models result in larger BH mass (several
%thousands solar mass) than the models Y and Yd where the rate in
%\cite{Y06} is used.

\section{Dependence on the DM annihilation rates}

A sufficiently high accretion rate of $\dot M \gsim 10^{-3}
M_{\odot} {\rm yr}^{-1}$ is required for Model Bd which results in
the most massive case in our studies (see Fig. 3). However, we
find that if the DM heating is stronger the evolution of the
accreting star can continue even with lower accretion rates. We
have tested this case by enhancing the DM capture rate by a factor
of 3. The same effect is expected if the DM annihilation
cross-section is enhanced by the Sommerfeld enhancement. Then the
steady growth and evolution were seen even for $\dot M \gsim
10^{-4} M_{\odot} {\rm yr}^{-1}$ (Model Cd). As long as the
accretion continues and the DM density is kept high enough, the
stellar mass will continuously increase and the star will become
brighter. These cases are shown by ($\times$3) to respective
models (e.g., Cd$\times$3, etc.) in Fig. 3. We note that these
models result in very massive stars, as large as $\sim
10^4M_{\odot}$ or larger (e.g., Model Cd$\times$3), and even for
models including radiative feedback where accretion rates decrease
with time (e.g., Model Fd$\times$3).

\section{Evolution toward gravitational collapse}

 We have shown that evolution of accreting 
population III stars may be stalled, 
or their lifetime will become essentially
infinite, if both the DM capture rate and the baryon mass
accretion rates are kept sufficiently high. For example, for the
fiducial DM parameter set, the evolution stalls if $\dot M \gsim
10^{-3}$ $M_\odot$ yr$^{-1}$. We find also that $\dot M \gsim
10^{-4}$ $M_\odot$ yr$^{-1}$ is enough for stalling if DM capture
rate $\propto \sigma \rho_\chi/m_\chi$ is enhanced by a factor of 3.

 Here we describe in detail how such stalling stars can evolve finally
to IMBHs.
There are three possible scenarios to end the stalling phase.
The first case occurs when ``nothing is changed''.
If mass accretion continues and the ambient DM density
stays constant, then the general relativistic instability is triggered
when the stellar mass is $M \sim 10^6 M_\odot$.
Then the star collapses to become a blackhole.

 It is, however, not certain whether or not, and at which mass the
instability actually occurs, because for a more massive star the
captured DM energy generation is larger. If the core temperature
of dark stars is too low, the general relativistic instability may
not take place even for $M > 10^6 M_\odot$. We will leave the
conclusion concerning this issue for our future work.

 The second case occurs when baryon mass accretion ceases
but the ambient DM density is unchanged. As shown in Fig.2 
when the stellar mass is less than about 10$^4
M_\odot$, the central temperature is about $T\sim 10^8$K. This
temperature is high enough to burn hydrogen. Therefore, if mass
accretion ceases when mass is below $\sim 10^4 M_\odot$, central
hydrogen is exhausted within $10^8$ years. On the other hand, a
star more massive than $10^5 M_\odot$ has a lower core
temperature, and thus cannot evolve quickly by merely stopping the
mass accretion because of strong DM heating.

 Even for $M \lsim 10^4 M_\odot$ dark stars, it is not trivial
to answer a question of whether the central temperature will rise
sufficiently high to proceed to helium and subsequent higher order
nuclear burning stages. We confirmed that for stars with $M \lsim
10^3 M_\odot$ the effect of DM annihilation is not very large.
These stars evolve up to the Fe-core formation, and the collapse
of the Fe-core is triggered as for stars without DM energy. DM
heating is stronger for $M \sim 10^4 M_\odot$ dark stars. We have
found that small amount of He-burning causes the stellar core to
expand, lowering the central temperature and density to stop the
He-burning, i.e., the stellar core oscillates. Eventually this
oscillating He-burning should be over (because stellar mass is
fixed now), but the lifetime will be much longer than the case
without DM heating. Therefore, this case may not necessarily
provide a prompt IBMH formation scenario.

 The third case occurs when the ambient DM density becomes lower.
This can happen, for example, when the star moves out of the
high-density region of DM. We find that in this case core collapse
is induced most efficiently. The evolution after the
central hydrogen exhaustion is particularly interesting for $M
\gsim 1.2\times 10^4 M_\odot$. 
To see this, we show in Fig. 4 the evolution after
the DM density is lowered by a factor of 0.3 when the stellar mass
is $M = 12,000M_\odot$. As shown in this figure,
the central temperature of the star increases when the DM density is
reduced. Then core hydrogen is exhausted within a few times $10^7$
years, and helium core burning follows.

\smallskip
\FIGURE{\epsfig{file=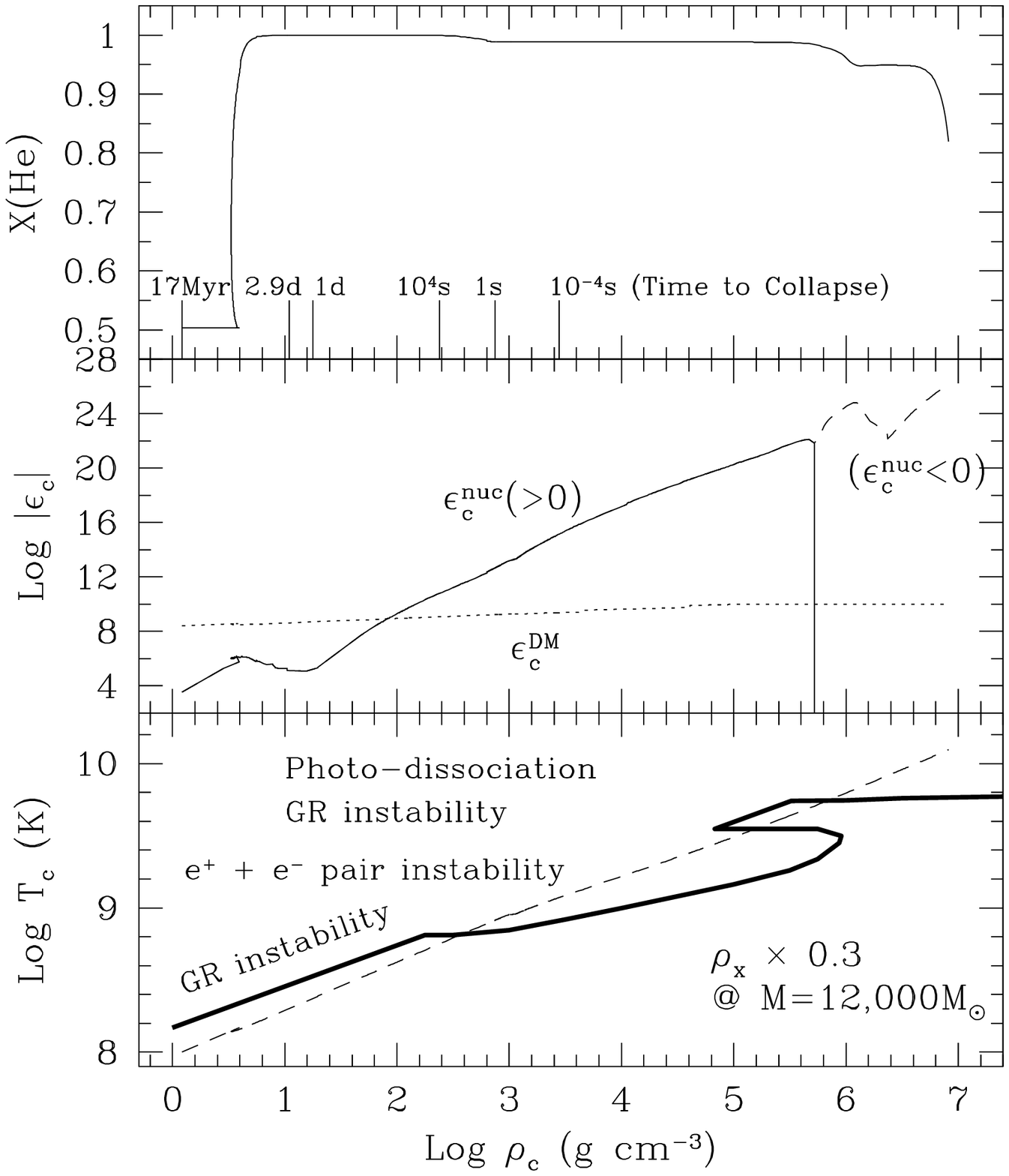,width=1\textwidth} 
        \caption{
The evolution of
Model Bd after the DM density is reduced by a factor of 0.3 when
the stellar mass is $M=12,000M_\odot$. The upper panel shows the
central He mass fraction as a function of central density.
Vertical lines with numbers indicate the time to collapse. 
At first, hydrogen burning takes place and helium mass fraction
increases from 0.5 to 1.0 in 17Myr. After the beginning of
He-burning the star collapses only in 3 days. 
%The collapse is too fast to burnout central He. 
The middle panel shows the energy
generation rate at the center by DM annihilation and nuclear
burning. The nuclear energy generation rate becomes negative (as
shown by the dashed line) after entering the (He)
photo-dissociation region. In the bottom panel the dashed line
shows the central density - temperature trajectory. The thick
solid line indicates that the star is unstable above this line.
}\label{fig4}}

\noindent
%As shown in this figure, soon after DM
%density is reduced the star starts to contract with increasing
%core temperature and density. Then central hydrogen is exhausted
%in 17Myr. The evolution after He-burning starts is even more rapid. 
As described in the figure caption,
after the He-ignition it takes only a few days before it
collapses to a BH. This is because pressure in the stellar
interior is dominated by radiation and thus the adiabatic index of
the equation of state, $\gamma$, is close to 4/3. The collapse
even more accelerates once the core enters the pair-instability
region at around Log$\rho_c\simeq 2.6$. Interestingly, the
collapse is so fast that the core directly collapse to a BH before
He exhaustion.

Since
the collapse is so rapid that there is not enough time to burnout
helium before the central temperature reaches $T \sim 10^{10}$K
where the composition is determined by the nuclear statistical
equilibrium. At this stage there is no way to stop the whole star
to collapse into a very massive BH.

% This ``direct-collapse'' of the He-core does not occur for a
%slightly less massive $M = 1.0\times 10^4 M_\odot$ star. The
%central density - temperature trajectory for the $M = 10^4
%M_\odot$ model is similar to the $M = 1.2\times 10^4 M_\odot$
%model but located slightly lower. Because of this small
%difference, the adiabatic index of the equation of state is
%slightly larger for the $M = 10^4 M_\odot$ model and thus collapse
%is decelerated by the energy generation from He-burning.
%Eventually, He-burning energy generation stops the core
%contraction and the star turns to expansion. Although this
%mechanism is similar to the pair-instability supernova occurring
%for a $M \sim 140-280 M_\odot$ star, the gravitational binding
%energy is huge for this very massive star and the whole star is
%never disrupted. After sometime, the core expansion should turn to
%contraction again. However with the DM energy, it takes longer
%than the usual stellar evolution. Because of this time-delay, IMBH
%formation from a $M = 10^4 M_\odot$ star might proceed slower than
%a $M = 1.2\times 10^4 M_\odot$ star.

\section{Discussions and summary}

 The accreting stars with DM heating were also explored in
\cite{Freese08,S09}. However, these authors adopted a very simple
polytropic model approach without carrying out actual stellar
evolution calculations. Their studies were also confined to
smaller accretion rates in \cite{Y06}, and only minimal
effects of DM capture were included. Consequently, we find that
our results are significantly different from theirs, even
qualitatively -- e.g., their mass does not go beyond $\sim$ 1000
$M_{\odot}$ even with DM.

 Spolyar et al. \cite{S09}
discussed that a dark star may have lower temperatures than a
normal star with the same mass, and so they may be distinguished.
However, this is in general quite difficult because the observed
temperature is determined by the location of the photosphere, that
may locate far above the star for an accreting star \cite{OP03}.
The geometry of the photosphere may not even be spherical if the
accretion is aspherical. In that case, only the reliable
observational data will be the total luminosity of the star, which
correlates with the stellar mass. 

 Since these very massive (dark) stars shine with the Eddington
luminosity (see Fig.3), they could 
actually be observed by future missions such as 
James Webb Space Telescope\footnote{http://www.jwst.nasa.gov},
Thirty Meter Telescope\footnote{http://www.tmt.org},
Giant Magellan Telescope\footnote{http://www.gmto.org},
and
European Extremely Large Telescope\footnote{http://www.eso.org/sci/facilities/eelt/}. 
In general, normal and dark
stars cannot be distinguished by the luminosity only.
For example, if DM has significantly smaller scattering cross section than
the fiducial value, $\sigma = 10^{-38}$cm$^{-2}$, then only adiabatic
contraction of dark matter
may contribute to the stellar evolution. In this case, according
to \cite{S09}, the DM effects will be small for $M > \sim
800M_\odot$. Hence, it will be quite difficult to distinguish
between normal and dark stars.
Nevertheless, if stars with Log $L/L_\odot \gsim 8.2$ or 
$M \gsim 4000M_\odot$ are discovered, it might be suggestive of
the effect of DM annihilation.

A direct collapse model for massive black formation was proposed 
by Bromm \& Loeb \cite{BL03}. In a proto-galactic size halo,
the gas can cool efficiently via hydrogen atomic cooling. 
In the absence of molecular hydrogen, due to a strong intergalactic
UV background, the gas can rapidly collapse to become a massive
blackhole. While the actual mass of the formed black hole is 
rather uncertain, the model offers a viable scenario for
the formation of massive blackholes without invoking
dark matter annihilation.
Interestingly, the model predicts that the gas 
mass accretion rate is very large in such a gas cloud \cite{SBH09}. 
Therefore, the combined effects of efficient gas cooling and 
the DM capture that we consider here, might make the formation
of massive blackholes even easier in large proto-galactic halos.

 In summary, we carried out {\it detailed stellar evolution
calculations of accreting dark stars} that include stellar
evolution consistently up to gravitational collapse. For the first
time we showed, {\it explicitly, that DM annihilation can make the
first stars much more massive than suggested so far}. These very
massive stars easily will gravitationally collapse to IMBHs of
mass as high as 10$^4 M_\odot$. Thus we offer a new possibility
that, formation of IMBHs with mass substantially higher than
1000$M_{\odot}$ in the early universe {\it opens a new viable path
for early formation of very massive SMBHs} with hierarchical
mergers.

\acknowledgments{ 
 We would like to thank M.~J. Rees, F. Takahashi and S. Mandal 
for useful comments 
and discussions. We thank an anonymous referee who gave 
many constructive comments, which improved the manuscript.
This work has been supported in part by the
grant-in-Aid for Scientific Research from the JSPS, 
MEXT of Japan, and
by the World Premier International Research Center
Initiative of MEXT.}


\begin{thebibliography}{999}
%\begin{thebibliography}{11}

%\bibitem{fltf}  Maths Dahlgren, {\it Package {\tt floatflt}, distributed 
%		with \LaTeXe{} 96/06/01}, 1994-1996.
\bibitem[1]{Fan}
X. Fan et al.,
{\it A Survey of $z >$5.7 Quasars in the Sloan Digital Sky Survey. II.
Discovery of Three Additional Quasars at $z > 6$},
\asj {\bf 125} {2003} {1649}.

\bibitem[2]{Willot}
C.~J. Willott, R.~J. McLure, M.~J. Jarvis,
{\it A $3\times 10^{9} M_{\odot}$ Black Hole in the Quasar SDSS J1148+5251 at $z=6.41$},
\apj {\bf 587} {2003} {L15}.

\bibitem[3]{Haiman}
Z. Haiman, A. Loeb,
{\it Cosmological Consequences of Population III Stars},
\apj {\bf 552} {2001} {459}.

\bibitem[4]{Madau}
P. Madau, M.~J. Rees,
{\it Massive Black Holes as Population III Remnants},
\apj {\bf 551} {2001} {L27 }.

\bibitem[5]{Volonteri}
M. Volonteri, M.~J. Rees,
{\it Quasars at $z=6$: The Survival of theFittest},
\apj {650} {2006} {669}.

\bibitem[6]{Li}
Y. Li et al.,
{\it Formation of $z\sim 6$ Quasars from Hierarchical Galaxy Mergers},
\apj {665} {2007} {187}.

\bibitem[7]{Tanaka}
T. Tanaka, Z. Haiman,
{\it The Assembly of Supermassive Black Holes at High Redshifts},
\apj {696} {2009} {1798}.

\bibitem[8]{BL03}
V. Bromm \& A. Loeb,
{\it Formation of the First Blackholes}, 
\apj {596} {2003} {34}.

\bibitem[9]{SBH09}
C. Shang, G. Bryan, Z. Haiman, 
{\it Supermassive Black Hole Formation by Direct Collapse: Keeping
Protogalactic Gas H$_2$--Free in Dark Matter Halos with Virial Temperatures
$T_{vir} \gsim 10^4$ K}, \arXivid{0906.4773} (2009).

\bibitem[10]{Bromm}
V. Bromm, N. Yoshida, L. Hernquist, C.~F. McKee,
{\it Formation of the First Stars and Galaxies},
\nature {458} {2009} {406}.

\bibitem[11]{Tegmark}
M. Tegmark {\it et al.},
{\it How Small Were the First Cosmological Objects ?},
\apj {474} {1997} {1}.

\bibitem[12]{Y03}
N. Yoshida, T. Abel, L. Hernquist, N. Sugiyama,
{\it Simulations of Early Structure Formation: Primordial Gas Clouds},
\apj {592} {2003} {645}.

\bibitem[13]{Bertone}
G. Bertone, D. Hooper, J. Silk,
{\it Partcile Dark Matter},
\prep {405} {2005} {279}.

\bibitem[14]{Ripamonti}
E. Ripamonti, M. Mapelli, A. Ferrara,
{\it The impact of dark matter decays and annihilations 
on the formation of the first structures},
\newjournal{MNRAS}{MNRAS}{375}{2007}{1399}.

\bibitem[15]{Spolyar}
D. Spolyar, K. Freese, P. Gondolo,
{\it Dark Matter and the First Stars: A New Phase of Stellar Evolution},
\prl {100} {2008} {1101}.

\bibitem[16]{Spergel}
W.~H. Press, D.~N. Spergel,
{\it Capture by the Sun of a Galactic Population of Weakly Interacting, 
Massive Particles},
\apj {296}{1985} {679}.

\bibitem[17]{Gould}
A. Gould,
{\it Resonant enhancements in weakly interacting massive particle 
capture by the earth},
\apj {321}{1987} {571}.

\bibitem[18]{Taoso}
M. Taoso, G. Bertone, G. Meynet, S. Ekstrom,
{\it Dark Matter annihilations in Pop III stars},
\prd {78}{2008}{123510}.

\bibitem[19]{Yoon}
S.-C. Yoon, F. Iocco, S. Akiyama,
{\it Evolution of the First Stars with Dark Matter Burning},
\apj {688}{2008} {L1}.

\bibitem[20]{S09}
D. Spolyar, P. Bodenheimer, K. Freese, P. Gondolo,
{\it Dark Stars: A New Look at the First Stars in the Universe},
\arXivid{0903.3701} (2009).

\bibitem[21]{Angle}
J. Angle et al.,
{\it Limits on Spin-Dependent WIMP-Nucleon Cross 
Sections from the XENON10 Experiment},
\prl {101}{2008}{091301}.

\bibitem[22]{Iocco}
F. Iocco,
{\it Dark Matter Capture and Annihilation on the First Stars: 
Preliminary Estimates},
\apj {677}{2008}{L1}.

\bibitem[23]{IoccoB}
F. Iocco, A. Bressan, E. Ripamonti, R. Schneider, A. Ferrara, P. Marigo, P.
{\it Dark matter annihilation effects on the first stars},
\newjournal{MNRAS}{MNRAS}{390}{2008}{1655}.

\bibitem[24]{Freese09}
K. Freese, P. Gondolo, J.~A. Sellwood, D. Spolyar,
{\it Dark Matter Densities During the Formation of the First Stars and 
in Dark Stars}, \apj {693}{2009}{1563}.

\bibitem[25]{Nata09}
A. Natarajan, J.~C. Tan, B.~W. O'Shea,
{\it Dark Matter Annihilation and Primordial Star Formation}, 
\apj {692}{2009}{574}.

\bibitem[26]{YOH08}
N. Yoshida, K. Omukai, L. Hernquist,
{\it Protostar Formation in the Early Universe}, 
\newjournal{Science}{Science}{321}{2008}{669}.

\bibitem[27]{Ohkubo06}
T. Ohkubo, H. Umeda., K. Maeda, K. Nomoto, T. Suzuki., S. Tsuruta, M.~J. Rees,
{\it Core-Collapse Very Massive Stars: Evolution, Explosion, and
Nucleosynthesis of Population III 500-1000 Msolar Stars},
\apj {645}{2006}{1352}.

\bibitem[28]{Ohkubo09}
T. Ohkubo, K. Nomoto, H. Umeda., N. Yoshida, S. Tsuruta,
{\it Evolution of Very Massive Population III Stars with Mass Accretion from
Pre-Main Sequence to Collapse},
\arXivid{0902.4573} (2009).

\bibitem[29]{OP03}
K. Omukai, F. Palla,
{\it Formation of the First Stars by Accretion},
\apj {589}{2003}{677}.

\bibitem[30]{ON98}
K. Omukai, R. Nishi,
{\it Formation of Primordial Protostars},
\apj {508}{1998}{1410}.

\bibitem[31]{MT08}
C.~F. McKee, J. Tan,
{\it The Formation of the First Stars. II. Radiative Feedback Processes and
Implications for the Initial Mass Function},
\apj {681}{2008}{771}.

\bibitem[32]{BL04}
V. Bromm, A. Loeb,
{\it Accretion onto a Primordial Protostar},
\newjournal{New Astronomy}{NewAstronomy}{353}{2004}{364}.

\bibitem[33]{Y06}
N. Yoshida, K. Omukai, L. Hernquist, T. Abel,
{\it Formation of Primordial Stars in a $\Lambda$CDM Universe},
\apj {613}{2006}{406}.

\bibitem[34]{Gao}
L. Gao, N. Yoshida, T. Abel, C.~S. Frenk, A. Jenkins, V. Springel,
{\it The first generation of stars in the $\Lambda$ cold dark matter cosmology},
\newjournal{MNRAS}{MNRAS}{378}{2007}{449}.

\bibitem[35]{MR}
V. Springel et al.,
{\it Simulations of the formation, evolution, and clustering 
of galaxies and quasars},
\nature {435}{2005}{629}.

\bibitem[36]{Freese08}
K. Freese, P. Bodenheimer, D. Spolyar, P. Gondolo,
{\it Stellar Structure of Dark Stars: A First Phase of Stellar Evolution 
Resulting from Dark Matter Annihilation},
\apj {685}{2008}{L101}.

%\bibitem{LC}    M. Goossens, F. Mittelbach, A. Samarin, 
%                {\it The \LaTeX{} Companion}, Addison-Wesley 1994.
%\bibitem{TeXbook} D. E. Knuth, {\it The \TeX book}, Addison-Wesley 1986.

\end{thebibliography}
\end{document}